\documentclass[aps,prl,preprint,superscriptaddress]{revtex4}

\usepackage{graphicx}
\usepackage{hyperref}
\usepackage{amsmath}

\begin{document}

\title{Coordinate transformation based design of confined metamaterial structures}

\author{Borislav Vasi\'c}
\email{bvasic@phy.bg.ac.yu}
\affiliation{Institute of Physics, Pregrevica 118, P. O. Box 68, 11080 Belgrade, Serbia}

\author{Goran Isi\'c}
\affiliation{School of Electronic and Electrical Engineering, University of Leeds, Leeds LS2 9JT, United Kingdom }
\affiliation{Institute of Physics, Pregrevica 118, P. O. Box 68, 11080 Belgrade, Serbia}

\author{Rado\v{s} Gaji\'c}
\affiliation{Institute of Physics, Pregrevica 118, P. O. Box 68, 11080 Belgrade, Serbia}

\author{Kurt Hingerl}
\affiliation{Zentrum f\"ur Oberfl\"achen- und Nanoanalytik und Universit\"at Linz, Altenbergerstr. 69, A-4040 Linz, Austria}

\begin{abstract}

 The coordinate transformation method is applied to bounded domains to design metamaterial devices for steering spatially confined electromagnetic fields. Both waveguide and free-space beam applications are considered as these are analogous within the present approach. In particular, we describe devices that bend the propagation direction and squeeze confined electromagnetic fields. Two approaches in non-magnetic realization of these structures are examined. The first is based on using a reduced set of material parameters, and the second on finding non-magnetic transformation media. It is shown that transverse-magnetic fields can be bent or squeezed to an arbitrary extent and without reflection using only dielectric structures.
\end{abstract}

\maketitle

\section{Introduction}

 The coordinate transformation method (CTM) \cite{pendry_science2006} employs the invariance of Maxwell equations under coordinate transformations \cite{post} to establish an equivalence between metric transformations and changes of material parameters \cite{leonhardt_generalrel}, \cite{leonhardt_geometryoflight}. The materials with parameters chosen to mimic a desired coordinate system are called the transformation media \cite{pendry_science2006}, \cite{schurig_raytracing2006}.

 CTM has been exploited for various computational problems such as: design of perfectly matched layers (PMLs) for simulations of open boundaries in finite-difference time-domain and finite element methods \cite{hugonin_pml}, \cite{shyroki_pml}, \cite{chew_pml}, \cite{teixeira_generalpml}, simplifying the geometry of complex computational domains \cite{chandezon}, \cite{ward1996}, \cite{ward_pendry_snom}, \cite{shyroki_efficient_modeling}, representing waveguide bends and twists by equivalent straight segments \cite{shyroki_bent}, \cite{shyroki_bent&twisted}, and formulating novel perturbation schemes for anisotropic materials \cite{kottke}. In \cite{pendry_science2006}, Pendry et al. suggested implementing transformation media as metamaterials and inspired the widespread use of CTM as an optical design tool.

 Several distinct strategies for the CTM-based design of metamaterial structures have been proposed. A continuous transformation of the whole space gives devices that are inherently invisible (like invisibility cloaks \cite{pendry_science2006}, electromagnetic field concentrators \cite{rahm_concentrator}, rotators \cite{chen&chan_rotator} and perfect lenses \cite{pendry&ramakrishna2003}, \cite{leonhardt_generalrel}). Various authors have previously applied CTM for design and modeling waveguides \cite{shyroki_bent}, \cite{shyroki_bent&twisted}, \cite{ozgun&kuzuoglu_miniaturization}, \cite{huangfu_bend}, \cite{donderici&teixeira_blueprintsforbends}. More recently, the embedded coordinate transformation method has been described \cite{rahm_embedded}, where transformation media are embedded into surrounding space yielding devices that transfer the transformed fields from the devices to their exterior.

 In this paper, we  apply CTM to domains bounded by transformation-invariant boundary conditions (BCs). Such an approach is inspired by the fact that in practice electromagnetic fields are always confined in space, like in waveguides and electromagnetic cavities or like electromagnetic beams.  A reflectionless CTM-based device can have a twofold function: it can rearrange at will the field distribution within itself and it can yield an orthogonally transformed (rotated) outgoing wave. The beam squeezer is considered as a typical example of the former and the waveguide bend as the typical example of the latter.

 The CTM-based design offers the possibility of realizing a device with a given function (e.g. cloaking) in countless ways but it is a rule of thumb that the resulting prescription for material parameters is very hard to implement. For optical applications, probably the biggest problem is to obtain a material with a controlled permeability. Non-magnetic CTM devices can be designed, but only for TM waves. Therefore, as our main goal for the considered structures (bend and squeezer) we choose achieving a non-magnetic realization. In Section III we show that using a reduced set of material parameters instead of the originally magnetic transformation media may be promising, provided the impedance-matching condition is reasonably satisfied. In Section IV we propose a different approach by considering transformations that yield non-magnetic transformation media. By giving examples of a non-magnetic waveguide bend and beam squeezer, we anticipate that in principle a structure with arbitrary shape can be designed as non-magnetic, provided that its optical length is unrestrained.

  The numerical simulations in the paper have been done using the COMSOL Multiphysics FEM solver.

\section{The coordinate transformation method and confined domains}

\begin{figure}[!htb]
\centering\includegraphics[width=7cm]{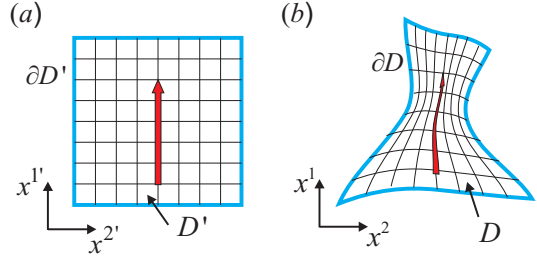}
\caption{(Color online) (a) The domain $D'$ in Cartesian coordinates $x^{i'}$, and (b) its image, the domain $D$ in curvilinear coordinates $x^i$. Blue lines stand for $\partial D$' and $\partial D$. Transformation media given by Eq. (\ref{eps_mu}) mimic the curvilinear coordinates $x^i$. Red arrows show the propagation directions of an electromagnetic wave, assuming homogeneous $\varepsilon^{i'j'}$ and $\mu^{i'j'}$.}
\label{fig_transformation}
\end{figure}

 Figure \ref{fig_transformation} depicts a domain in (a) Cartesian coordinates $x^{i'}$ ($D'$) and (b) curvilinear coordinates $x^i$ ($D$) given by
  \begin{equation} \label{coord_transf}
x^i=x^i(x^{1'},x^{2'},x^{3'}), \quad \Lambda^i_{i'}=\frac{\partial x^i}{\partial x^{i'}},
\quad \Lambda^{i'}_{i}=\frac{\partial x^{i'}}{\partial x^i}.
\end{equation}
   Domain boundaries are denoted by $\partial D'$ and $\partial D$ with BCs corresponding to either zero-field or perfect electric conductor (PEC), both of which are invariant under coordinate transformations. The electric permittivity and magnetic permeability of the medium within $D'$ are labeled with $\varepsilon^{i'j'}$ and $\mu^{i'j'}$, respectively.

The Maxwell equations are invariant under coordinate transformations \cite{post}, whereas the material parameters in $D$ read \cite{leonhardt_geometryoflight}, \cite{rahm_concentrator}
\begin{equation} \label{eps_mu}
\varepsilon^{ij}=\Lambda^{-1} \Lambda^i_{i'} \Lambda^j_{j'} \varepsilon^{i'j'}, \quad
\mu^{ij}=\Lambda^{-1} \Lambda^i_{i'} \Lambda^j_{j'} \mu^{i'j'},
\end{equation}
where $\Lambda$ is the determinant of $\Lambda^i_{i'}$ and summation over repeated indices (running from $1$ to $3$) is assumed. The electric and magnetic fields (one-forms) in $D'$ are $E_{i'}$ and $H_{i'}$, so the fields in $D$ read
\begin{equation} \label{fields_change}
E_i=\Lambda^{i'}_i E_{i'}, \quad H_i=\Lambda^{i'}_i H_{i'}.
\end{equation}
The above coordinate transformation can be physically implemented by interpreting $x^i$ as Cartesian coordinates and choosing the permittivity and permeability of the physical media within $D$ (the transformation media) to coincide with $\varepsilon^{ij}$ and $\mu^{ij}$ \cite{leonhardt_geometryoflight}.

\begin{figure}[!htb]
\centering\includegraphics[width=7cm]{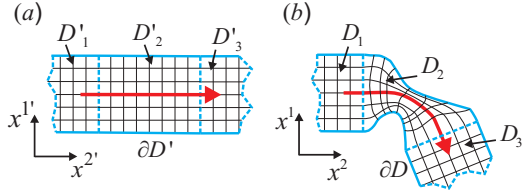}
\caption{(Color online) Explaining how transformed fields are obtained outside transformation media. Dashed boundaries indicate that only relevant parts are shown while domains may extend arbitrarily beyond them. For simplicity, the media are assumed to be homogeneous in $D'$. (a) An electromagnetic wave (red arrow) propagates along a straight line in $D'$. (b) The wave is transformed by transformation media in $D_2$ maintaining the orientation after passing through.}
\label{fig_subdomains}
\end{figure}

Figure \ref{fig_subdomains} explains how transformed fields outside transformation media
are obtained. The domain $D'$ is divided into three subdomains, $D'_1$, $D'_2$, $D'_3$ and transformed to $D_1$, $D_2$, $D_3$, respectively. Within $D'_1$ and $D'_3$ it is assumed that the transformation satisfies
 \begin{equation} \label{orthogonal}
\Lambda^{i'}_i \Lambda^{j'}_j  \delta^{ij}=\delta^{i'j'}, \quad \Lambda=1, \quad
x^{i'} \in D'_1 \cup D'_3.
\end{equation}
 Equation (\ref{orthogonal}) means that $D_1$ and $D_3$ have the same shape as $D'_1$ and $D'_3$, respectively.  Subdomain $D'_2$ is transformed to $D_2$ so that it continuously connects $D_1$ and $D_3$. Now Eqs. (\ref{eps_mu}) and (\ref{orthogonal}) assert that $\varepsilon^{ij}$ and $\mu^{ij}$ for $x^i \in D_1 \cup D_3$ are same as the corresponding $\varepsilon^{i'j'}$ and $\mu^{i'j'}$ for $x^{i'} \in D'_1 \cup D'_3$ except that they may be rotated (together with $D'_1$ and/or $D'_3$). Since a continuous transformation (\ref{coord_transf}) is established between entire $D'$ and $D$, the fields are transformed according to (\ref{fields_change}) in $D_1$ and $D_3$ in spite of the fact that media remained unchanged with respect to that in $D'_1$ and $D'_3$. This way a recipe is established for a reflectionless rotation of field distribution and polarization or change of propagation direction of a confined electromagnetic field that passes through a domain ($D_2$) with transformation media.

 The method of transferring modified fields away from transformation media described above is an alternative to the one based on embedded coordinate transformations by Rahm et al. \cite{rahm_embedded}, \cite{rahm_beam_bend}. The benefit here is that the absence of reflection comes as a simple consequence of transformation continuity.

 Another application of the concept of bounded domains is in the design of structures with a desired external shape without affecting the way fields perceive the internal geometry. A typical example is in waveguide miniaturization \cite{ozgun&kuzuoglu_miniaturization} or squeezing an electromagnetic beam, which we describe in detail below.

  The performance of any CTM-based device is, in principle, limited only by physical properties of the metamaterials implementing the transformation media. In the general case, parameters required for transformation media are such that causality implies the occurrence of both dispersion and absorption over any finite frequency band.  These limitations have been reported for cloaking devices in \cite{pendry_science2006}. Moreover, the fabrication is complicated since the metamaterial will need to be both anisotropic and with spatially varying properties in general case, as can be seen from (\ref{eps_mu}). Still, perhaps the biggest obstacle in optical applications is the difficulty in engineering the magnetic response. For these reasons, in CTM applications it is crucial to consider how can the material requirements be released or simplified even if it is sometimes achieved at the expense of departing from ideally required values.

   In the following two sections we consider two characteristic devices, the waveguide bend and beam squeezer, and discuss their implementation with respect to the above mentioned problems. Our main aim is to remove magnetism completely and that can be done for the TM polarized waves. In section III this is achieved by substituting the original (magnetic) transformation media parameters with a reduced non-magnetic set having the same dispersion relation. In section IV we describe transformations that yield a non-magnetic transformation media.

\section{Device design and implementation with a reduced set of material parameters}

\subsection{Waveguide bend}

\begin{figure}[!htb]
 \centering\includegraphics[width=7cm]{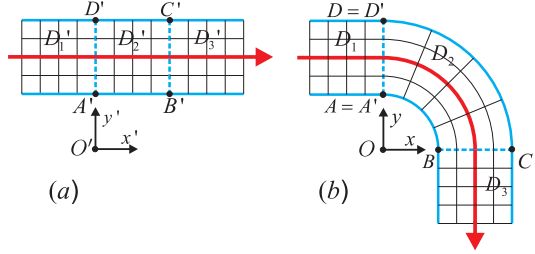}
 \caption{(Color online) (a) Domain $D'$ from Fig. \ref{fig_subdomains} (a) and (b) the corresponding domain $D$. The rectangular subdomain $D^{'}_{2}$ is transformed to the annular segment shaped subdomain $D_2$. The red arrows show the wave propagation direction when a metamaterial with parameters (\ref{par_rotator}) is placed inside $D_2$.}
 \label{fig_rotator}
 \end{figure}

   The waveguide/beam bend shown in Fig. \ref{fig_rotator} has previously been treated in \cite{donderici&teixeira_blueprintsforbends}, \cite{huangfu_bend}, \cite{rahm_beam_bend}, \cite{jiang_bending}, \cite{vasic_ps}. The underlying transformation is
    \begin{equation} \label{coord_rotator}
 x=y'\sin\left(\kappa x'\right), \quad y=y'\cos\left(\kappa x'\right).
 \end{equation}
 In this and all the following examples, we assume $z=z'$ and free-space background $\varepsilon^{i'j'}=\varepsilon_0 \delta^{i'j'}$ and $\mu^{i'j'}=\mu_0 \delta^{i'j'}$. By $\varepsilon^{ij}=\mu^{ij}$ we will label the relative permittivity and permeability of the transformation media.

   The rectangle $A'B'C'D'$ is transformed to the annular segment $ABCD$. $L=\vert A'B'\vert$ is the optical length of the bend while $R_1=\vert OA \vert$ and $R_2=\vert OD \vert$ are its inner and outer radius. If the bend angle is $\alpha$, then $\kappa L=\alpha$. The relative permittivity is diagonal in cylindrical coordinates $(r,\phi,z)$ and reads
 \begin{equation}\label{par_rotator}
  \varepsilon^{\alpha\beta}=\mathrm{diag} \left( (\kappa r)^{-1}, \kappa r, (\kappa r)^{-1} \right), \quad \alpha, \beta=r, \phi, z.
  \end{equation}

  From now on we are concerned only with TM waves ($z$ being the magnetic axis) for which only $\varepsilon^{r r}$, $\varepsilon^{\phi \phi}$ and $\mu^{z z}=\varepsilon^{z z}$ are relevant. Following the procedure used in \cite{schurig_science2006} and \cite{cai_shalaev_cloak} for the cloak, a reduced parameter set for the bend from Fig. \ref{fig_rotator} (b) is found as \cite{jiang_bending}, \cite{vasic_ps}

 \begin{equation}\label{par_rotator_reduced}
 \varepsilon^{rr}=(\kappa r)^{-2}, \quad \varepsilon^{\phi \phi}=1, \quad \mu^{zz}=1.
 \end{equation}

 In both the ideal (\ref{par_rotator}) and reduced set of parameters (\ref{par_rotator_reduced}), the free parameter $\kappa$ appears. Since parameters (\ref{par_rotator}) are impedance-matched to vacuum irrespective of the value of $\kappa$, it affects only the phase shift in passing through the bend. However, reduced parameters $(\ref{par_rotator_reduced})$ are impedance-matched to vacuum only for $\kappa r=1$, so the value of $\kappa$ determines the amount of reflection at the entrance $AD$ and exit $BC$ of the bend. The value of $\kappa$ that minimizes reflection depends on the incoming wave but for a general case, it is reasonable to set
  \begin{equation}\label{matching_condition}
 \kappa=\frac{1}{R_C}, \quad R_C=\frac{R_1+R_2}{2},
 \end{equation}
 matching the impedance along the central line of the bend. Now we see that in case of reduced parameters, the bend geometry ($R_1$ and $R_2$) will play the deciding role in its performance through impedance mismatch away from $r=R_C$.

\begin{figure}[!htb]
 \centering\includegraphics[width=7cm]{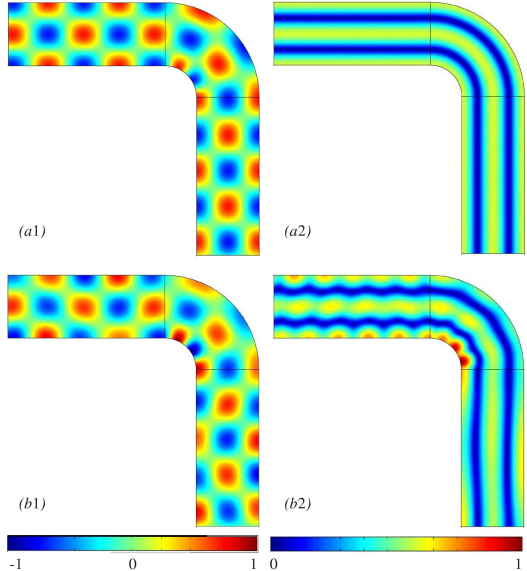}
 \caption{(Color online) FEM simulation results for the $2\mathrm{GHz}$ $\mathrm{TM}_2$ mode excited at the left edge of the structures: (a) the waveguide bend with ideal and (b) with reduced set of the material parameters. On the left side, (x1), the real part of magnetic field phasor is shown and on the right side, (x2), the magnetic field magnitude.  The bend angle is $90\,^{\circ}$, $R_1=0.1m$, $R_2=0.3m$ and $\kappa=1/0.2m$ in both cases.}
 \label{fig_rotator_fem}
 \end{figure}

 Figure \ref{fig_rotator_fem} shows numerical simulations results for a $90^{\circ}$ waveguide bend. The second TM mode, $\mathrm{TM}_2$, with frequency $2\mathrm{GHz}$ is excited at the left edge of structures. The simulated domains have PEC boundaries and are terminated by PMLs to simulate infinite waveguides. For reference, Fig. \ref{fig_rotator_fem} (a1) and (a2) show the case where ideal parameters (\ref{par_rotator}) have been used. The lower part of Fig. \ref{fig_rotator_fem}, (b1) and (b2), shows the case of reduced parameters (\ref{par_rotator_reduced}) together with (\ref{matching_condition}). From the weak standing wave pattern in front of and the dominantly $\mathrm{TM}_2$ pattern behind the bend, Fig. \ref{fig_rotator_fem} (b2), we see that reflection and modal mixing are low, so the device works fine.

 A possible implementation of the non-magnetic bend with reduced set of parameters is as a structure comprised of many concentric annular layers with homogeneous and isotropic permittivities. The continuous variation of $\varepsilon^{r r}$ given by (\ref{par_rotator_reduced}) is first approximated by $N$ annular layers  with constant $\varepsilon^{r r} =(\kappa r_i)^{-2}$ and thickness $d$, $r_i$ being the inner radius of the layer and $i=1,2,...,N$. To obtain the anisotropic permittivity, each of the $N$ layers is further divided into $n$ layers with homogeneous permittivities $\varepsilon_k$ and thickness $\delta_k d$, while $\sum^n_{k=1} \delta_k=1$ ($\delta_k$ are the relative thicknesses), as in \cite{huang_layered_cloak}. $n$ is usually taken to be 2, but we consider the case of $n=3$, as well.

  If the wavelength is large compared to $d$, each of $N$ layers can be considered to be an anisotropic medium with effective dielectric permittivity \cite{wood_subwavelength_imaging_layers}, \cite{huang_layered_cloak}
  \begin{equation} \label{layered_effective_medium2}
  \frac{1}{\varepsilon^{rr}}=\sum_{k=1}^n \frac{\delta_k}{\varepsilon_k}, \quad
  \varepsilon^{\phi \phi}=\sum_{k=1}^n \delta_k \varepsilon_k.
  \end{equation}

 First we consider the case $n=2$ and take that $\delta_1=\delta_2=0.5$. From (\ref{par_rotator_reduced}) we find $\varepsilon_1$ and $\varepsilon_2$ as
 \begin{equation}\label{dielectric_layers_permittivity}
 \varepsilon_1=1-\sqrt{1-(\kappa r_i)^{-2}}, \quad \varepsilon_2=1+\sqrt{1-(\kappa r_i)^{-2}}.
 \end{equation}
 If  $1-(\kappa r_i)^{-2}$ is to be positive within the bend (this corresponds to passive/lossless dielectrics), the condition (\ref{matching_condition}) cannot be met. We have found that this is the main obstacle in realizing better bends with dielectric layers.

\begin{figure}[!htb]
 \centering\includegraphics[width=7cm]{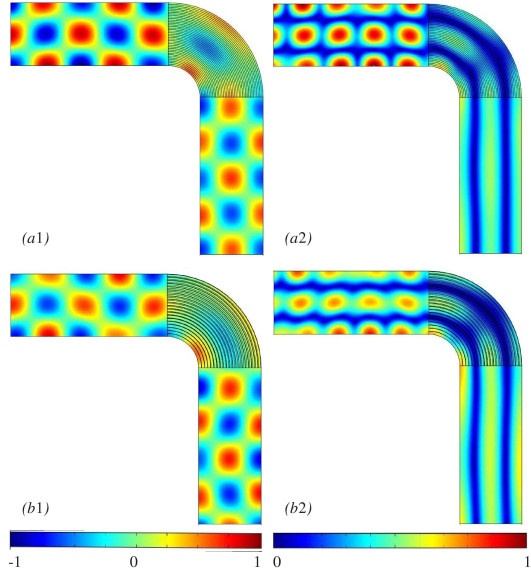}
 \caption{(Color online) FEM simulation results for the $2\mathrm{GHz}$ $\mathrm{TM}_2$ mode excited at the left edge of structures. (a) The waveguide bend with $\kappa=1/0.1m$, $N=20$ and $40$ dielectric layers. Layer permittivities are gradually changed according to (\ref{dielectric_layers_permittivity}). (b) The waveguide bend with $\kappa=1/0.08m$, $N=20$ and $60$ dielectric layers. Layer permittivities are fixed ($\varepsilon_1=0.8$, $\varepsilon_2=0.05$, and $\varepsilon_3=8$) while their thicknesses vary with $r$. (x1) show the real part of the magnetic field phasor while (x2) are the magnetic field magnitudes. }
 \label{layered_bend}
 \end{figure}

 Figure \ref{layered_bend} (a1) and (a2) show the simulation results for $\mathrm{TM}_2$ mode passing through a layered bend with $n=2$ where $\kappa=1/R_1$ ($R_1=0.1m$) and $40$ layers ($N=20$) have been used. Layer permittivities vary from $0.06$ to $1.94$. That the bending effect works very well is seen from a negligible modal mixing, as both the reflected and transmitted wave consist of almost only $\mathrm{TM}_2$. However, compared to the case shown in Figure \ref{fig_rotator_fem} (b1) and (b2), there is a significant reflection (indicated by the pronounced standing wave in front of the bend) and the transmitted power is reduced. As we previously discussed, this is due to the choice of $\kappa$.

 To improve on the implementation with $n=2$, we consider the case of $n=3$ anticipating that it will allow us to achieve spatially varying effective parameters (\ref{par_rotator_reduced}) by varying $\delta_k$ while keeping $\varepsilon_k$ fixed. The realization with fixed permittivities is clearly more favorable from a practical standpoint since it is easier to control the thickness of layers than their permittivity.

 Taking $\varepsilon_k$ as fixed parameters, from (\ref{layered_effective_medium2}) and $\sum_k \delta_k=1$, we find $\delta_k$ as
 \begin{equation} \label{delta_system}
 \begin{bmatrix}
 \delta_1 \\ \delta_2 \\ \delta_3
 \end{bmatrix}=
 \begin{bmatrix}
 1 & 1 & 1 \\
 \varepsilon_1 & \varepsilon_2 & \varepsilon_3 \\
 \varepsilon_1^{-1} & \varepsilon_2^{-1} & \varepsilon_3^{-1}
 \end{bmatrix}^{-1}
 \begin{bmatrix}
 1\\ 1\\ (\kappa r)^2
 \end{bmatrix}.
 \end{equation}
 The solution for $\delta_k$ for a given value of $\kappa r$ has to satisfy
 \begin{equation} \label{positive_delta}
 \delta_k >0, \quad  k=1,2,3,
 \end{equation}
 because $\delta_k$ represent the relative thicknesses so they must be positive. Equation (\ref{positive_delta}) determines which values of $\kappa$ can be obtained. An additional problem arises from the fact that $\varepsilon_k$ should also be positive. We have found that in case of negative $\varepsilon_k$ the effective medium description (\ref{layered_effective_medium2}) is ruined for propagation parallel to the layers both due to the extinction of waves in layers with negative permittivity and due to the excitation of surface states on their interfaces.

  It is straightforward to show that (\ref{delta_system}), (\ref{positive_delta}) and $\varepsilon_k>0$ can be simultaneously met only if $\kappa r\geq1$. With (\ref{dielectric_layers_permittivity}) we have seen that (\ref{matching_condition}) cannot be met in the case of $n=2$, so now we see that it is not possible for $n=3$ either. The same conclusion applies for any $n$.

 Simulation results for the bend with $N=20$ and $60$ layers are shown in Fig. \ref{layered_bend} (b1) and (b2). The value of $\kappa$ is chosen to be $1/0.08\mathrm{m}$. Layer permittivities used in simulation are $\varepsilon_1=0.8$, $\varepsilon_2=0.05$ and $\varepsilon_3=8$. The reflection is still higher than in Fig. \ref{fig_rotator_fem} (b1) and (b2), but lower than in the case of the $n=2$ bend. The improvement over the $n=2$ is probably due to using more layers. The main advantage of the $n=3$ bend is, as noted before, in it being implemented using only three different homogeneous and isotropic materials.

 We have, so far, shown that this bend (the one discussed in literature so far) can, for TM polarization, be implemented using non-magnetic materials so that modal mixing is negligible, but with evident reflection depending on details of the TM wave. In Section IV we will further improve on this by considering a non-magnetic transformation media (for a different bend) that will remove reflection completely.

 \subsection{Beam squeezer}

\begin{figure}[!htb]
 \centering\includegraphics[width=7cm]{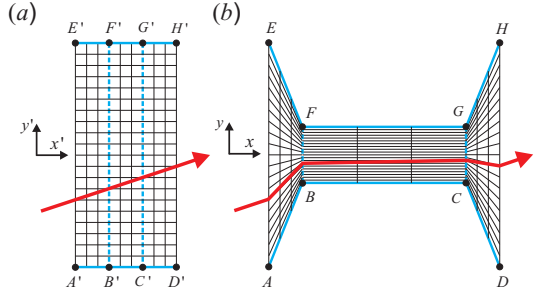}
 \caption{(Color online) Transformation underlying the beam squeezer. (a) Rectangular domain in Cartesian coordinates (a), and (b) the squeezed domain in curvilinear coordinates $x^i$. The transformation media  for the trapezoidal subdomains are (\ref{par_wormhole1}) and (\ref{par_wormhole2}) for the middle subdomain. The red arrows illustrate the beam propagation direction corresponding to the simulation in Fig. \ref{fig_wormhole_fem}.}
 \label{fig_wormhole}
 \end{figure}

 Here we use ideas from Section II to design a device for manipulating free-space electromagnetic beams. Figure \ref{fig_wormhole} shows the underlying transformation for a beam squeezing structure. The beam is compressed within the left trapezoidal subdomain, then it passes through the long channel, and finally, it is expanded to the initial width within the right trapezoidal subdomain. Vacuum is assumed as the exterior of the device. Similar devices have independently been considered in \cite{ozgun&kuzuoglu_miniaturization} and \cite{zhang_narrow_slit} except that stretching of the channel has not been investigated there.

  While the bend is a typical example of a CTM-based structure used for obtaining a rotated outgoing wave, here the main aim is to make a structure that complies with external requirements (width and length) leaving the outgoing wave unaffected. According to Fig. \ref{fig_wormhole}, two rectangles, the left (L), $A'B'F'E'$, and the right one (R), $C'D'H'G'$, are gradually narrowed down from $\vert AE \vert =\vert A'E'\vert=w_0$ to $\vert BF\vert=w$ and connected by the rectangle between them (the channel) which has been narrowed down to $w$ and stretched out from $\vert B'C'\vert=l_0$ to $\vert BC \vert =l$. The side rectangles are transformed using
  \begin{equation}\label{coord_trapezoid}
  x=x'+x_{\alpha}, \quad y-Y_0=\Delta_{\alpha}(y'-Y_0),\quad
  \Delta_{\alpha}= a_{\alpha}x+b_{\alpha},\quad \alpha=R,L.
  \end{equation}
  The introduced parameters are defined with $a_L=(w-w_0)w_0^{-1}s'^{-1}$, $b_L=1-a_L x_{A}$, $a_R=-a_L$, $b_R=1-a_R x_{D}$, $x_L=0$, $x_R=l-l_0$, $s'=\vert A'B'\vert=\vert C'D'\vert=x_{B}-x_{A}=x_{D}-x_{C}$, $y=Y_0$ is the symmetry axis of trapezoidal subdomains (in this case $Y_0=0$) while $x_{P}$ represents the $x$ coordinate of the point denoted by $P$. The material parameters for the left and the right trapezoids are found as
  \begin{equation} \label{par_wormhole1}
  \varepsilon^{ij}_{\alpha}=\mu^{i j}_{\alpha}=
  \begin{bmatrix}
  \Delta_{\alpha}^{-1} & a_{\alpha} \Delta_{\alpha}^{-2} (y-Y_0) & 0\\
  a_{\alpha} \Delta_{\alpha}^{-2} (y-Y_0) & a_{\alpha}^2 \Delta_{\alpha}^{-3} (y-Y_0)^2+\Delta_{\alpha}&0\\
  0&0&\Delta_{\alpha}^{-1}
  \end{bmatrix}.
  \end{equation}
  The transformation of the rectangle $B'C'G'F'$ to the rectangle $BCGF$ is a simple rescaling
  \begin{equation}\label{coord_rect}
  x=(x'-x_{B'})\xi+x_{B}, \quad (y-Y_0)=\gamma (y'-Y_0) ,
  \end{equation}
  where $\xi=ll_0^{-1}$ and $\gamma=ww_0^{-1}$ are the stretching and compression parameters. The material parameters for the $BCGF$ rectangle are
  \begin{equation} \label{par_wormhole2}
  \varepsilon^{ij}=\mu^{i j}=\mathrm{diag} (\xi\gamma^{-1}, \xi^{-1}\gamma, \xi^{-1}\gamma^{-1}).
  \end{equation}

\begin{figure}[!htb]
\centering\includegraphics[width=7cm]{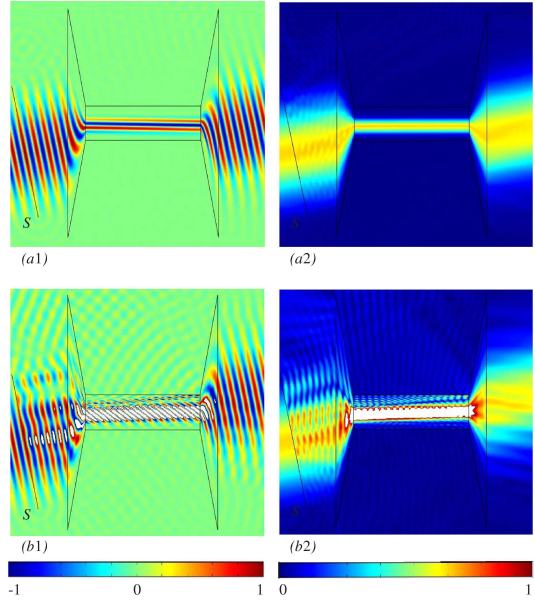}
\caption{(Color online) FEM simulation of the beam squeezer: (a) ideal parameters ($\gamma=0.15$ and $\xi=39$) and (b) reduced parameters ($\gamma=0.15$ and $\xi=1$). (x1) Real part of the magnetic field phasor, (x2) magnetic field magnitude distribution. $S$ is the source of the TM beam. The color maps are saturated for better contrast.}
\label{fig_wormhole_fem}
\end{figure}

  The effect of geometrical transformations on electromagnetic fields can be discussed on the example of the channel in Fig. \ref{fig_wormhole_fem}. Using (\ref{coord_rect}) and (\ref{fields_change}) the field intensities in the channel are found to be
 \begin{equation} \label{field_changes_rectangular}
 E_x=\xi^{-1}E_{x'}, \quad E_y=\gamma^{-1}E_{y'}, \quad H_z=H_{z'},
 \end{equation}
 with $H_z$ unaltered since the $z$-axis is left invariant. This shows that the compression (stretching) along the $y$-direction ($x$-direction) is accompanied by increased (decreased) field component intensity along that direction. The squeezing structure focuses the fields concentrating all the power flow to an arbitrarily small (not limited by diffraction) cross section. It could be used for various sensor applications where a strong signal improves the sensitivity or for imaging purposes.

  Figure \ref{fig_wormhole_fem} (a1) and (a2) show the simulation of a Gaussian TM beam (source denoted by $S$) with oblique incidence passing through the squeezer with ideal material parameters.  To better illustrate how the fields propagate through the structure, the compression and the stretching were set to moderate values, $\gamma=0.15$ and $\xi=39$. The structure entrance is twice as wide as the beam, ensuring that the beam does not reach structure boundaries. In this way, the zero field BC is satisfied.  Having passed through the narrow channel, the beam leaves the squeezer translated (due to stretching of the channel) along the $x$-axis retaining its initial propagation direction and the field distribution. Since the structure is theoretically perfect the slight reflection is only an artefact of numerical simulation and can be decreased with higher mesh density. However, it also indicates an increasing sensitivity to material parameter deviations (here occurring due to numerical discretization) with higher compression (small $\gamma$) or stretching (bigger $\xi$).

  The reduced set of material parameters for the squeezer is obtained by setting \mbox{$\mu^{z z}_{red}=1$} while changing the relevant components of permittivity so that the dispersion relation remains unaffected \cite{schurig_science2006}:
 \begin{equation}\label{reduced_parameters_wormhole}
 \varepsilon^{xx}_{red}=\mu^{zz}\varepsilon^{xx}, \quad
 \varepsilon^{xy}_{red}=\mu^{zz}\varepsilon^{xy}, \quad
 \varepsilon^{yy}_{red}=\mu^{zz}\varepsilon^{yy}.
 \end{equation}
 From our experience with the bend, we expect that reflection may be significantly decreased by impedance-matching to vacuum along the central line  of the device ($y=Y_0$). As in the case of the bend, we find the impedance-matching condition to be that the optical length equals the physical length, i.e. $\xi=1$.

 Simulation results for the squeezer with reduced material parameters and $\xi=1$ are given in Fig.  \ref{fig_wormhole_fem} (b1) and (b2). As can be seen, the device retains its function but with a visible distortion of the transmitted beam. Also, appreciable reflection appears from the entrance and exit of the device, as indicated by standing wave patterns in front and within the device in Fig. \ref{fig_wormhole_fem} (b2). To highlight the overall field distribution, color maps have been saturated (regions with pronounced standing wave peaks have been cut off). In simulated cases with $\xi \neq 1$ (not shown) we have found strong reflection and obscured field pattern. Thus the proposed squeezer can be realized with reduced parameters, but it is crucial to have its impedance along the central line matched to the exterior.

\section{Devices based on non-magnetic transformation media}

 In the previous section we have attempted to substitute the ideal (magnetic) transformation media parameters with a non-magnetic reduced set having the same dispersion $\omega(\mathbf{k})$. That approach yielded good results but there was always the problem of reflection due to impedance mismatch.

 As in the previous section a free-space background is assumed, $\varepsilon_0^{-1} \varepsilon^{i'j'}=\mu_0^{-1} \mu^{i'j'}=\delta^{i'j'}$. As long as the transformation leaves the $z$-direction invariant, $z=z'$, one of the optical axes of the transformation media is parallel to it and we have
 \begin{equation} \label{mu_z}
 \mu^{zz}=\Lambda^{-1},
 \end{equation}
 which is the only relevant permeability component for TM waves. Thus if
 \begin{equation}\label{condition_non-magnetic}
 \Lambda=1,
 \end{equation}
 a non-magnetic transformation media for TM waves is obtained. If we agree to evaluate the volume in $x^i$ coordinates as if they were Cartesian (see note after (\ref{fields_change}) in Section II), the condition (\ref{condition_non-magnetic}) means that every subdomain of $D'$ is mapped to a subdomain of $D$ with the same volume.

 So far, this idea received little attention in the literature. Chen and Chan consider layered dielectric structures that act as non-magnetic transformation media in \cite{chen_layered_systems}. In \cite{luo_cloak_axial_inv_parameters}, Luo et al. describe a cylindrical cloak with spatially invariant permeability $\mu^{zz}$  which is based on a similar idea, except that $\Lambda=\mathrm{const}>1$ because the cloaking transformation decreases the volume of a domain by creating a hole in it.

  We now show that the waveguide bend and beam squeezer can be designed with a transformation that satisfies (\ref{condition_non-magnetic}) and thus conceptually solve the problem of bending and squeezing TM beams using non-magnetic structures.

 \subsection{Non-magnetic waveguide bend}

  \begin{figure}[!htb]
\centering\includegraphics[width=7cm]{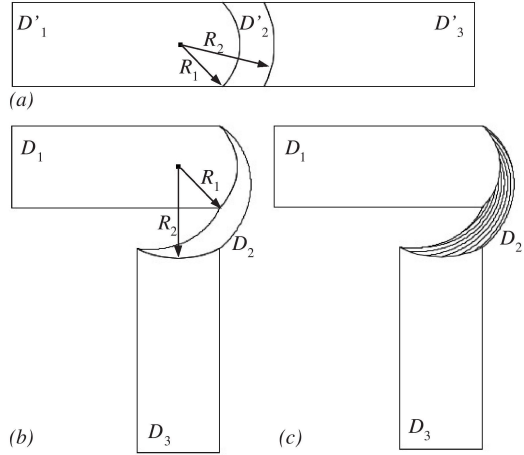}
\caption{(Color online) (a) Straight waveguide structure is transformed to (b) a bent waveguide. Transformation of $D'_2$ to $D_2$ is given by (\ref{non-magnetic_bend_transf}). (c) Curved layers within $D_2$ for implementation using dielectric layers.}
\label{fig_non-magnetic_bend_transf}
\end{figure}

 The TM beam shifter described in \cite{rahm_embedded} is non-magnetic and (\ref{condition_non-magnetic}) is satisfied. The shifter does not change a field propagation direction, it only translates the field. A change of a field propagation direction can be achieved by a shift of the polar angle in cylindrical coordinates. This is shown in Fig. \ref{fig_non-magnetic_bend_transf} where the annular ring segment $D'_2$ in Fig. \ref{fig_non-magnetic_bend_transf} (a) is transformed to the subdomain $D_2$ of the bent waveguide in Fig. \ref{fig_non-magnetic_bend_transf} (b). The transformation reads
 \begin{equation}\label{non-magnetic_bend_transf}
 r=r', \quad \theta=\theta'+\theta_0\frac{f(r')-f(R_1)}{f(R_2)-f(R_1)}, \quad z=z',
 \end{equation}
 where $\theta_0$ is the bend angle, $f(r')$ an arbitrary continuous function of $r'$,  $R_1$ and $R_2$ are the inner and outer radius of $D'_2$. The relative permittivity tensor $\varepsilon^\alpha_\beta$ in cylindrical coordinates ($\alpha,\beta=r,\phi,z$) is given by
 \begin{equation}\label{non-magnetic_bend_parameters}
 \varepsilon^{\alpha \beta}=
 \begin{bmatrix}
 1 & mr & 0\\
 mr & 1+mr^2 & 0 \\
 0 & 0 & 1
 \end{bmatrix}, \quad
 m=\frac{\theta_0}{f(R_2)-f(R_1)}\frac{\partial f(r')}{\partial r'}
 \end{equation}

  The transformation (\ref{non-magnetic_bend_transf}) is the same as the one used in \cite{chen&chan_rotator} for the field rotator. The bend is obtained by cutting out a segment from the field rotator. This device can be implemented using curved isotropic and homogeneous dielectric layers when $f(r')=\ln (r')$ \cite{chen_layered_systems}. It is shown in Fig. \ref{fig_non-magnetic_bend_transf} (c) where alternating dielectric layers with permittivities $\varepsilon_1=0.037$ and $\varepsilon_2=27.33$ are assumed.

  \begin{figure}[!htb]
\centering\includegraphics[width=7cm]{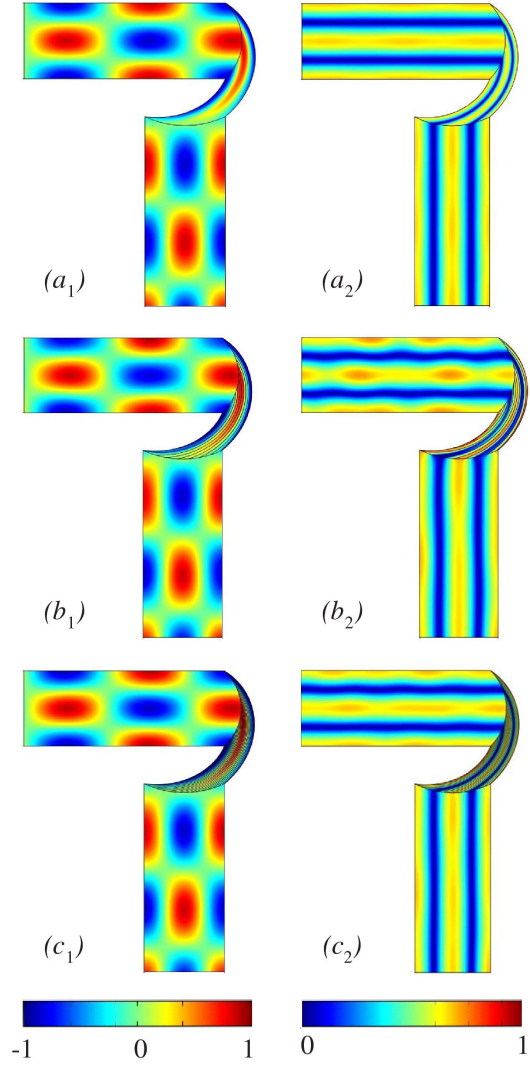}
\caption{(Color online) FEM simulation results for non-magnetic waveguide bend: (a) ideal bend, (b) bend with $13$ dielectric layers and (c) bend with $26$ dielectric layers. Layer permittivities are constant, $\varepsilon_1=0.037$ and $\varepsilon_2=27.33$. (x1) show the real part of the magnetic field phasor, while (x2) are the field magnitude distributions.}
\label{fig_non-magnetic_bend_fem}
\end{figure}

  Figure \ref{fig_non-magnetic_bend_fem} (a) shows the simulation results for the non-magnetic waveguide bend with PEC boundaries. The $1.66\mathrm{GHz}$ $\mathrm{TM}_2$ wave is excited on the left edge. The structure is theoretically ideal so the weak reflection is a numerical error that can be decreased by increasing the mesh density or by increasing the ratio $R_2/R_1$ so that the bending is less abrupt. Figure \ref{fig_non-magnetic_bend_fem} (b) shows the simulation for a bend realized using $13$ dielectric layers. There is a slightly higher reflection than in Fig. \ref{fig_non-magnetic_bend_fem} (a). The bend in Fig. \ref{fig_non-magnetic_bend_fem} (c) is realized with $26$ layers giving practically the same results as the ideal structure from Fig. \ref{fig_non-magnetic_bend_fem} (a).

  These results demonstrate that a reflectionless and non-magnetic waveguide bend for TM waves can be fabricated using only isotropic and homogeneous dielectric layers. Compared to the bend with reduced parameters analyzed in Section III, the non-magnetic bend clearly shows a superior performance that comes at the expense of a more complicated geometry.

 \subsection{Non-magnetic beam squeezer}

   \begin{figure}[!htb]
\centering\includegraphics[width=7cm]{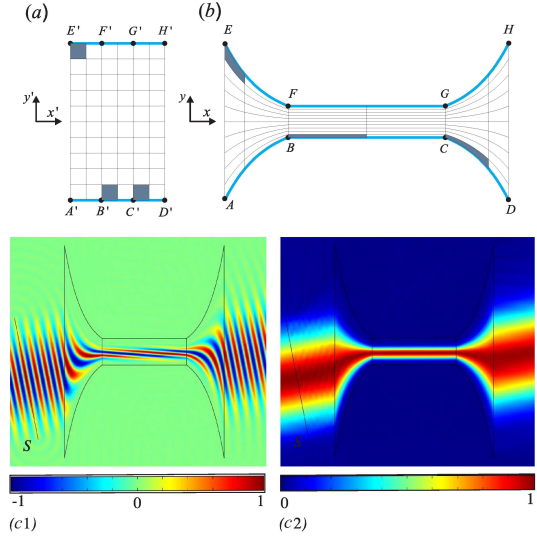}
\caption{(Color online) (a) Rectangular domain in $x^{i'}$ transformed to (b) the domain in $x^i$ by transformation given in (\ref{non-magnetic_wormhole_transf1}) and (\ref{non-magnetic_wormhole_transf2}). Few subdomains in (a) and (b) are shaded to emphasize that their areas are equal since $\Lambda=1$. Simulation of a 7.5GHz TM Gaussian beam excited at $S$: (c1) Real part of the magnetic field phasor distribution, (c2) magnetic field amplitude distribution. Simulation parameters: $\gamma=0.125$, $s'=0.06\mathrm{m}$, $\vert BF \vert =0.1\mathrm{m}$ and $l=0.32\mathrm{m}$. }
\label{fig_squeezer}
\end{figure}

 Figure \ref{fig_squeezer} (a) and (b) show the transformation used for the non-magnetic beam squeezer. The shaded rectangles in Fig. \ref{fig_squeezer} (a) and corresponding shaded subdomains in Fig. \ref{fig_squeezer} (b) having the same area, illustrate the point that $\Lambda=1$. For rectangles $A'B'F'E'$ ($L$) and $C'D'H'G'$ ($R$) the transformation is given by
 \begin{equation} \label{non-magnetic_wormhole_transf1}
 x=\frac{1}{a_\alpha}\ln \left| x'+\frac{b_\alpha}{a_\alpha}\right|+c_\alpha, \quad
 y=(a_\alpha x' +b)y', \quad \alpha=R, L,
\end{equation}
with parameters $a_R=-a_L=(1-\gamma)/s'$, $s'=\vert A'B' \vert$, $\gamma=\vert BF \vert /\vert B'F'$, $b_L=1-a_Lx'_{A'}$, $b_R=1-a_Rx'_{D'}$, $l=\vert BC \vert$ and
\begin{equation}
 c_L=x'_{A'}-\frac{1}{a_L}\ln\left(\frac{s'}{1-\gamma}\right), \quad
 c_R=c_L+l-\frac{2s'}{1-\gamma}\ln\left(\frac{\gamma s'}{1-\gamma}\right).
\end{equation}

 The relative permittivity tensor of transformation media within $ABEF$ and $CDHG$ reads
\begin{equation}
 \varepsilon^{ij}_\alpha=
 \begin{bmatrix}
 f^{-2} &  a_\alpha yf^{-2} & 0\\
 a_\alpha yf^{-2} & a_\alpha^2y^2f^{-2}+f^2 & 0\\
 0 & 0 & 1
 \end{bmatrix}
 \end{equation}
 with $f_\alpha=\vert a_\alpha \vert \exp\left(a_\alpha(x-c_\alpha)\right)$.

 The transformation of the rectangle $B'C'G'F'$ to the rectangle $BCGF$ is determined by the condition that the compression $\gamma$ is reciprocal to the stretching $\xi=\vert BC \vert/\vert B'C' \vert$:
\begin{equation} \label{non-magnetic_wormhole_transf2}
 x=\gamma^{-1} x'+\frac{\mathrm{ln}\gamma}{a_L}-\frac{(1-\gamma)x_{A'}+s'}{\gamma},
  \quad y=\gamma y',
\end{equation}
  so the parameters of media in the rectangle $BCGF$ are
  \begin{equation}
  \varepsilon^{ij}=\mathrm{diag}(\gamma^{-2},\gamma^2,1).
  \end{equation}

 The simulation of a TM Gaussian beam passing through the squeezer in Figure \ref{fig_squeezer} (c1) and (c2) confirms the perfect performance of the proposed device.

 This completes the demonstration of non-magnetic realization of the two principal examples in manipulating confined TM fields. The approach we have taken in this section is clearly better than the one with reduced material parameters that is widely used for the design of non-magnetic CTM-based devices. It might not be apparent immediately, but the condition $\Lambda=1$ is quite lax and is equivalent to fixing the phase of the transformed field, which we had to do even in the case of reduced material parameters to minimize reflection (recall fixing $\kappa$ for the bend and $\xi$ for the squeezer).

 In summary, the application of CTM to spatially confined electromagnetic fields was considered.
 It was used to give various solutions for the waveguide bend and beam squeezer. Both approximate and exact non-magnetic realizations were found. In the case of structures with a reduced set of material parameters, the influence of impedance mismatch and its dependence on free parameters were investigated. Transformations with unit Jacobian determinant were recognized as promising for the design of non-magnetic devices and used to find novel non-magnetic solutions for the bend and squeezer. Several implementations using isotropic and homogeneous layered systems were considered. On the example of the bend with reduced parameters, it was shown that the inhomogeneity of effective permittivity tensor can be successfully controlled by varying only the layer thicknesses. The implementation of the non-magnetic bend with such layers was shown to be practically perfect. Its application is in optical waveguides that allow TM polarized waves. The non-magnetic squeezer is interesting for focusing and imaging applications since it is not limited by diffraction.

\section*{Acknowledgments}
 This work is supported by the Serbian Ministry of Science project 141047. G. I. acknowledges support from ORSAS in the \mbox{U. K.} and the University of Leeds. R. G. acknowledges support from EU FP7 project Nanocharm. K. H. is grateful to the Austrian NIL-meta-NILAustria project from FFG for partial support. We are, also, grateful to Photeon and Heinz Syringer from Photeon Technologies for financial support and Johann Messner from the Linz Supercomputer Center for technical support.

\end{document}